\documentstyle [a4,epsfig]{article} 
\newcommand{\bc}{\begin{center}} 
\newcommand{\ec}{\end{center}} 
\newcommand{\be}{\begin{equation}} 
\newcommand{\ee}{\end{equation}} 
\newcommand{\bea}{\begin{eqnarray}} 
\newcommand{\eea}{\end{eqnarray}} 
\newcommand{\bs}{\begin{subequations}} 
\newcommand{\es}{\end{subequations}} 
\newcommand{\beq}{\begin{eqalignno}} 
\newcommand{\eeq}{\end{eqalignno}} 

\begin{document}

\thispagestyle{empty} 
 
\vspace{2.0cm}

\bc

\huge{Dissipation and memory domains in the quantum model of brain} 
 
\vspace{1.2cm} 
 
\large{Eleonora Alfinito and Giuseppe Vitiello} \\  
\small 
{\it Dipartimento di Fisica, Universit\`a di Salerno, 84100  
Salerno, Italy\\ 
INFM Sezione di Salerno and 
INFN Gruppo Collegato di Salerno\\ 
alfinito@sa.infn.it\\ 
vitiello@sa.infn.it }

\vspace{1.5cm}

\ec

\normalsize 
 
{\bf Abstract} 
 
We shortly review the dissipative quantum model of brain and its parametric  
extension.

\vspace{4.5cm}

\newpage

\section{Introduction}

Statistical mechanics predicts macroscopic laws exhibiting ordering and  
regularities in the behavior of systems made by a large number of
components. 
Schr\"odinger, however, points out that such "regularities  
only in the average" of statistical origin are not  
enough to explain the high stability and the high degree of ordering of  
living matter. Pretending to explain the biological stability in  
terms of the regularities of statistical origin would be the "classical  
physicist's expectation", that "far from being trivial, is wrong"  
\cite{SRO}. Schr\"odinger makes the distinction between  
ordering generated by the "statistical mechanisms" and ordering generated  
by "dynamical" quantum (necessarily quantum!) interactions among the atoms  
and the molecules. Such a distinction between the "two ways of producing  
orderliness" is of crucial relevance in the study of living  
matter and of brain. In such a line of thought has to be framed the quantum  
model of brain originally proposed by Ricciardi and Umezawa in 1967  
\cite{UR} and further developed by Stuart, Takahashi and Umezawa \cite{S1,   
S2} (see also \cite{CH}). To a similar framework also belongs the dynamical  
model for living matter based on the boson condensation mechanism, proposed  
by H. Fr\"ohlich \cite{FR} in the middle of the 1960s. The Fr\"ohlich model  
was further developed in the 1980s thus leading to the quantum field theory  
(QFT) approach to living matter \cite{DG, PRL}.  
 
In the 1960s, the theory of the dynamical generation of macroscopic ordered  
states in many body physics was developed and experimentally tested in the  
observations on superconductors, ferromagnets, superfluids, crystals. Such  
a theory is based on the key mechanism of the spontaneous breakdown of  
symmetry in QFT by which long range correlations (the Nambu-Goldstone (NG)  
boson modes) are dynamically generated \cite{AND, U}. On the other hand, it  
was already experimentally well established, since Lashley's and Pribram's  
\cite{P2,PR} pioneering work, that many functional activities of the brain  
involve extended assembly of neurons. On this basis, Pribram introduced  
concepts of Quantum Optics, such as holography, in brain modeling  
\cite{P2}. In the brain, information is indeed  
observed to be spatially uniform. While  
the activity of the single neuron is experimentally observed in form  
of discrete and stochastic pulse trains and point processes, the  
``macroscopic'' activity of large assembly of neurons appears to be  
spatially coherent and highly structured in phase and amplitude  
\cite{fre2, fre3}. The quantum model of brain is firmly founded on such an  
experimental evidence. The model is in fact primarily aimed to the  
description of non-locality of brain functions, especially of memory  
storing and recalling. The mathematical  
formalism in which the model is formulated is the one of QFT. The  
motivation for using such a formalism is explained by Umezawa in one of his  
last papers \cite{UC}: "In any material in condensed matter physics any  
particular information is  
carried by certain ordered pattern    
maintained by certain long range correlation mediated by massless    
quanta. It looked to me that this is the only way to memorize    
some information; memory is a printed pattern of order supported    
by long range correlations...If I could know    
what kind of correlation, I would be able to write down the    
Hamiltonian, bringing the brain science to the level of condensed    
matter physics." In the model the  
"dynamical variables" are not the neurons and the other cells 
(Stuart, Takahashi and Umezawa have observed \cite{S1}, with a  
pleasant sense of humor, that  
"it is difficult to consider neurons as quantum objects"), but  
they are
identified \cite{YA1} with 
those of the electrical dipole vibrational  
field of the water  
molecules \cite{PRL} and of other biomolecules present in the brain  
structures, and with the ones  
of the associated NG modes, named the dipole wave quanta (dwq)  
\cite{VT}.   
The model exhibits interesting features related with  
the r\^ole of microtubules in the brain  
activity \cite{P2,PR,YA} and its 
extension to dissipative dynamics \cite{VT} allows a huge memory  
capacity. The dissipative  
quantum model has been investigated \cite{PV} also in  
relation with the modeling of neural networks exhibiting long range  
correlation among the net units. The parametric extension of the  
model has been also considered \cite{brain2000}.  
 
In the following we briefly review some aspects of  
the dissipative parametric quantum model. For sake of shortness we omit  
mathematical details. These are reported in the quoted literature.

\section{The quantum model of brain and dissipation} 
 
In QFT spontaneous breakdown of symmetry occurs when the  
dynamical equations are invariant  
under some group, say $G$, of continuous transformations, but the  
minimum energy state (the ground state or vacuum) of the system is not  
invariant under the full group $G$. In such a case the vacuum is an ordered  
state and massless particles (the NG bosons), also called  
collective modes, are dynamically generated and acting as {\em long range  
correlations} \cite{AND, U}. Propagating over the whole system, they are  
the carriers of the ordering information: {\it order manifests itself as a  
global property which is 
dynamically generated}. For example, the magnetic order in ferromagnets is  
a diffused, i.e. macroscopic, feature of the system. In this way in QFT it  
is possible to describe "macroscopic quantum systems".  
 
Since the collective mode is a massless particle, its presence ({\it  
condensation}) in the vacuum does not add energy to it: the stability of  
the ordering is thus insured. As a further consequence, infinitely many  
vacua with different degrees of order may exist, corresponding to different 
densities of the condensate. In the infinite  
volume limit they are each other physically (unitarily) inequivalent and  
thus they represent possible physical phases of the system. The actual  
phase in which the system sits is determined by some external agent, acting  
as a trigger of the symmetry breakdown.  
The observable specifying the ordered state, called order parameter, acts  
as a macroscopic variable since the collective modes manifest a  {\em  
coherent} dynamical behavior.  
The order parameter is specific of the kind of symmetry into play and its  
value may be considered as a {\em code} specifying the vacuum.  
 
The conclusion is that  
stable long range correlation and diffuse, nonlocal properties  
related with a code specifying the 
system state are dynamical features of {\em quantum} origin. 
 
We remark that the von Neumann theorem in quantum mechanics (QM) states  
that all the representations of the canonical commutation relations  
are unitary (and therefore physically) equivalent in  
systems with a finite number of degrees of freedom. In QFT, where the  
number of degrees of freedom  
is infinite, the von Neumann theorem does not hold and   
infinitely many unitarily inequivalent representations exist. It is because  
of the existence of such inequivalent representations that dynamically  
generated ordering may exist in a stable state in QFT.  
 
In the quantum model of brain \cite{UR} memory recording is 
represented by the ordering induced in the ground  
state by the 
condensation of NG modes dynamically generated through the  
breakdown of the rotational  
symmetry of the electrical dipoles of the water molecules. They are the  
dipole wave quanta (dwq). The trigger of the symmetry breakdown is the  
external informational input. The "code" classifying the  
recorded information is identified with the "order parameter".  
The recall mechanism is described as the  
excitation of dwq from the ground state under the action of an  
external imput similar to the one which has previously produced the memory  
recording. 
 
The high stability of memory demands that the long range correlation modes  
(the dwq) must be in the lowest energy state (the   
ground state), which also guarantees that memory is easily   
created and readily excited in the recall process.  
The long range correlation must also be quite   
robust in order to survive against the constant state of   
electrochemical excitation of the brain and the continual response  
to external stimulation. It can  
be shown \cite{PRL} that the time scale associated with the  
coherent interaction in the QFT of electrical dipole fields for water  
molecules is of the order of $10^{-14}$ $sec$, thus much shorter than  
times associated with short range interactions, and therefore these  
effects are well protected against thermal fluctuations.  
At the same time, the brain electrochemical activity must be coupled to the  
correlation modes. It is indeed the electrochemical activity observed by  
neurophysiology that provides a first response to external stimuli. The  
brain is then modeled \cite{S1, S2} as a "mixed" system involving two  
separate but interacting levels. The memory level is a  
(macroscopic) quantum dynamical level, the electrochemical activity is at a  
classical level.  
 
Note that vacua labeled by different code numbers are accessible  
only through a sequence of phase transitions form one to another one of  
them. This process destroys previously stored  
information ({\em overprinting}). This problem of {\em memory capacity}  
arises because in the model there is only one kind of code number since  
only one kind of symmetry is assumed (the dipole rotational symmetry). In  
order to avoid overprinting a huge  
number of symmetries (a huge number of codes) \cite{S1} could be assumed.  
This would introduce serious difficulties and spoil the model practical  
use.  
 
However, the memory  
capacity can be enormously enlarged \cite{VT} by considering the intrinsic  
dissipative character of the brain dynamics (the brain is an {\it open  
system}  
continuously coupled to the environment). In fact, let us denote the dwq  
variables by $A_{k}$ and recall that  
the canonical formalism for dissipative systems \cite{QD, VT} requires the  
introduction of a ``mirror'' set of  
dynamical variables, say ${\tilde A}_{k}$, describing the environment. The  
number ${\cal{N}}_{A}$ for all $k$ of  
$A_{k}$-modes, condensed in the vacuum ${|0>}_{\cal N}$, 
constitutes the ``code'' of the information. The crucial point of  
dissipative dynamics is that 
the vacuum state is now defined to be the state in which the  
{\em difference} ${\cal{N}}_{A} -{\cal{N}}_{\tilde A}$ for all $k$ is zero.    
Thus we see that there are infinitely many vacua, each one  
corresponding to a different value of the code ${\cal{N}}_{A}$. Moreover,  
each of these states is unitarily inequivalent to the other ones, and thus  
``protected'' from unwanted interferences with other memory states. 
The "brain  (ground) state" is then represented as 
the collection (or the  superposition) of the full set of memory  
states ${|0>}_{\cal N}$, for all  $\cal N$. 
  
The brain is thus described as a complex system with a  
huge number of macroscopic quantum states (the memory states). The   
dissipative dynamics introduces 
$\cal N$-coded "replicas" of the system and information printing  
can be  performed in each replica without reciprocal destructive  
interference. A huge  
memory capacity is thus achieved \cite{VT}.

\section{The parametric extension of the dissipative model}

In the parametric dissipative model the dwq frequency is assumed to be  
time-dependent \cite{brain2000}. Remarkably, it is found that in such a  
case the couple of equations describing the dwq $A$ and the ``doubled''  
modes ${\tilde A}$ is equivalent to the spherical Bessel equation of  
order $n$ ($n$ integer or zero) \cite{brain2000}.  
 
The coupled system $A-{\tilde A}$ is then found to be described by a  
parametric oscillator of frequency $\Omega_{n}(k,t)$. The time-dependence  
of this frequency means that  
energy is not conserved in time and therefore that the $A-{\tilde A}$  
system does not constitute a "closed" system. However, when $n  
\rightarrow \infty$, $\Omega_n$ approaches to a constant, i.e.  
energy is conserved and   
the $A-{\tilde A}$ system gets "closed" in such a  
limit. Thus, as $n \rightarrow \infty$  
the possibilities of the system $A$ to  
couple to ${\tilde A}$ (the environment) are "saturated": the system  
$A$ gets {\it fully} coupled to $\tilde A$. This suggests that $n$  
represents the number of {\it links} between $A$ and ${\tilde A}$.  
When $n$ is not very large (infinity), the system $A$ (the brain) has  
not fulfilled its capability to establish links with the external  
world. We also have that $n$ ``graduates'' the rate of variations in time  
of the frequency, i.e. the "rapidity" of the system  
response to  external stimuli.  
 
In order that memory recording may occur, $\Omega_{n}(k,t)$ has to be real.  
Such a condition is satisfied only in a limited span of 
time $T_{k,n}$. For fixed $k$, $T_{k,n}$ grows linearly in $n$, which means  
that the time span  
useful for memory recording (the ability of memory  
storing) grows  as the number of links with the external world grows: more  
the system is "open" to the external world (more are the links), better  
it can memorize (high ability of learning).  
 
The reality condition also implies that only modes with $k$ greater or  
equal to a threshold $\tilde k(n,t)$ can be recorded. Such a kind of  
"sensibility" to external stimuli also depends on some characteristic  
parameter $L$ of the system. 
 
This intrinsic infrared cut-off in turn implies that only wave-lengths 
$\lambda \le {\tilde \lambda} \propto  
{1 \over {{\tilde k}(n,t)}}$ are allowed: thus (coherent)  
domains of sizes less  
or equal to ${\tilde \lambda}$ are involved in the memory  
recording. Such a cut-off shrinks in time for  
a given $n$. On the other hand, a growth of $n$ opposes to such a  
shrinking. These cut-off changes correspondingly reflect on the  
memory domain sizes. 
It is thus expected that, for given $n$, "more impressive" is  
the external stimulus, i.e. greater is the number of high $k$ momentum  
excitations produced in the brain, more "focused" is the "locus" of the  
memory.  
 
 
\begin{figure}[t] 
\caption{``Lives'' of $k$ modes, for fixed $n$} 
\epsfig{file=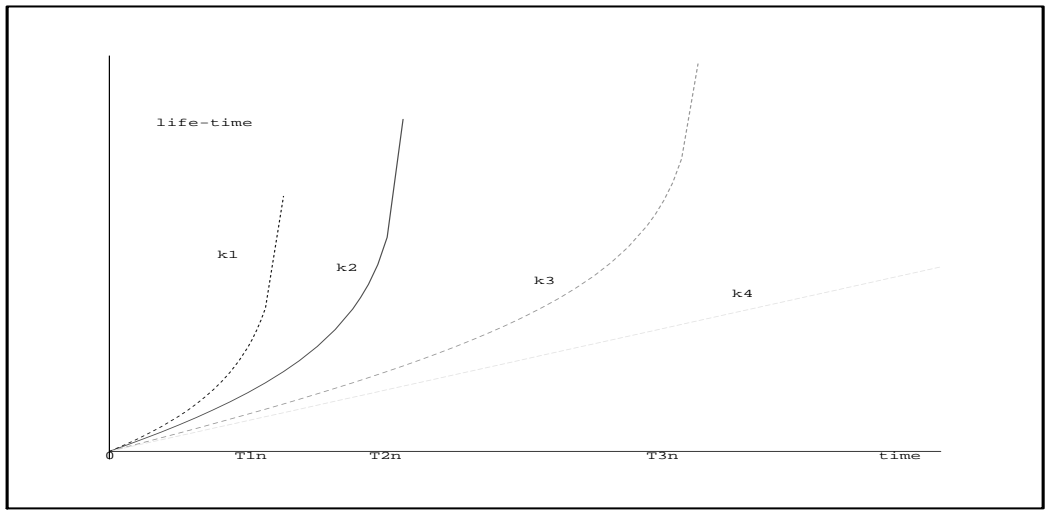,width=0.5\linewidth,height=1.0\linewidth,angle=-90}   
\end{figure}


\begin{figure}[t] 
\caption{``Lives'' of $k$ modes, for growing $n$ and fixed $k$} 
\epsfig{file=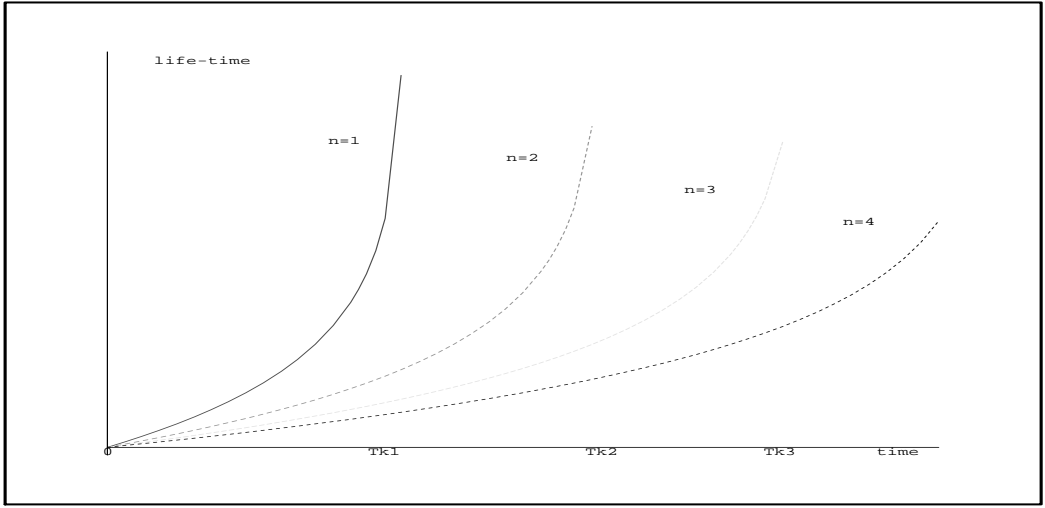,width=0.5\linewidth,height=1.0\linewidth,angle=-90} 
\end{figure}


The finiteness of the size of the domains implies that transitions through  
different vacuum states at given $t$ 
become possible. As a consequence, both the phenomena of association  
of memories and of confusion of memories, which would be avoided in  
the regime of strict unitary inequivalence among vacua (i.e. in the  
infinitely long wave-length regime), are possible \cite{VT}. 
 
We also find that modes with larger $k$ have a "longer" life with reference  
to time $t$. Only modes satisfying the reality condition are  
present at certain time  
$t$, being the other ones decayed (Fig. 1 and 2). 
This introduces an hierarchical organization of memories  
depending on their life-time: memories with a specific spectrum of  
$k$ mode components  
may coexist, some of them "dying" before, some other ones persisting  
longer. The sizes of the associated memory domains  
are correspondingly larger or smaller.

\section{Conclusions}

The results presented above appear to fit qualitatively well with the  
physiological observations \cite{GRE} that more the brain  
relates to external environment, more neuronal connections will form.  
We are referring to functional or effective connectivity, as opposed to the  
structural or anatomical one. While the last one can be  
described as quasi-stationary, the  
former one is highly dynamic with modulation time-scales  
in the range of hundreds of milliseconds \cite{GRE}. These functional  
connections may quickly change and new  
configurations of connections may be formed extending over a domain  
including a larger or a smaller number of neurons. 
Such a picture finds a possible  
description in our model, where the coherent domain formation, size  
and life-time depend on  
the number of links that the brain sets with its environment and   
on internal parameters.

The finiteness of the domain size implies a non-zero effective  
mass of the dwq. These therefore propagate with a greater ``inertia'' than  
in the case of infinite volume where they are massless. The domain  
correlations are consequently established with a certain time-delay.  
This appears to agree  
with the physiological observation that the recruitment of  
neurons in a correlated assembly is achieved with a certain delay  
after the external stimulus  
action \cite{LI, GRE}. The dwq effective non-zero mass also acts  
as a threshold in the excitation energy of dwq so that, in order to  
trigger the recall process   
an energy supply equal or greater than such a 
threshold is required. When the energy supply is lower than threshold a  
"difficulty in recalling" may be experienced. At the same 
time, however, the threshold may positively act as a "protection" 
against unwanted perturbations (including thermalization) and 
cooperate to the stability of the memory  state.  In the case of zero 
threshold (infinite size domain) any replication signal could excite   
the recalling  and the brain would fall in a state of "continuous  
flow of memories" \cite{VT}.  
 
Finally, note that {\it after} information has been  
recorded, memory stability implies that the brain cannot be brought to the  
state in which it was {\it before} the information printing occurred ({\it  
Now}, you know it!...). Thus, the same fact of getting information introduces  
a partition in  
the time evolution, it introduces the {\it distinction} between the past 
and the  
future, a distinction which did not exist {\it before} the information  
recording (the psychological {\it the arrow of time}).
 
In fact, the main feature of dissipative quantization is that at each 
time $t$ the  
system ground state  $|0(t)>$ is labeled by $t$ (``foliation''), so that at  
$t' \neq t$ the ground state $|0(t')>$ is unitary inequivalent to  
$|0(t)>$: in its time evolution the system runs over unitarily  
inequivalent representations. Such a non-unitary  
time evolution is found to be controlled by the entropy variation rate,  
as expected since dissipation implies irreversibility  
\cite{QD}. The psychological {\it arrow of time} thus naturally emerges in  
the dissipative quantum model. Moreover, the system ground state is found  
to be a thermal state \cite{U, QD, VT}, and the psychological arrow of time  
is actually concord \cite{brain2000} with the thermodynamical arrow of time  
and with the cosmological arrow of time (defined by the expanding Universe  
direction) \cite{double, Hawking}.

\end{document}